\documentclass[prl,preprint,longbibliography]{revtex4-1}
\usepackage{graphicx}
\usepackage{bbm}
\usepackage{bm}
\usepackage{amsmath}
\usepackage{amssymb}
\usepackage{color}
\usepackage[pdftex,colorlinks]{hyperref}
\usepackage[english]{babel}

\hypersetup{urlcolor=blue}
\def\U{\ensuremath{\mathcal{U}}}
\def\P{\ensuremath{\mathcal{P}}}
\def\id{\ensuremath{\mathbbm{1}}}
\DeclareMathOperator{\Tr}{Tr}
\begin{document}
\title{Modified Kubo formula with a complex force term for weak measurement}
\author{Antonio \surname{Di Lorenzo}}
\email{dilorenzo@infis.ufu.br}
\affiliation{Universidade Federal de Uberl\^{a}ndia, Uberl\^{a}ndia, MG, Brazil}
\affiliation{CNR-IMM-UOS Catania (Universit\`a), Consiglio Nazionale delle Ricerche,
Via Santa Sofia 64, 95123 Catania, Italy}

\begin{abstract}
In a seminal work, Aharonov, Albert, and Vaidman showed that by having a weak interaction 
between a system and a detecting apparatus, the average output of the latter could be much larger than the maximum eigenvalue of the observed quantity (times the amplification factor). 
 This does not always happen, however: the observed system must subsequently undergo a second measurement, on the output of which the result of the first one is conditioned. This procedure is known as postselection.
On the other hand, linear response theory describes how the observables of a quantum system change upon 
perturbation by a weak classical external force. In a measurement, the measured system applies a generalized 
force to the measuring apparatus, leading to an observable change in the latter. 
It appears natural, then, to unify the treatment of weak measurements with an extended version of linear response theory 
that accounts for a force introduced by an external quantum system. 
Here, we show how the postselection introduces a complex force term and we provide a modified Kubo formula working 
in the non-linear regime. 
\end{abstract}
\maketitle
\emph{What is a measurement?\textemdash}
Some may have a very restricted idea of a measurement, limiting the concept to the case when a single observation 
on a system allows to infer total information about a dynamical quantity of interest in a second system that interacted with the 
first in a controlled way. 
For us, instead, a measurement is any act of inference about a system done by observing another system (which can be called indifferently the detector, the meter, the probe, the apparatus, or the ancilla)  
that has interacted with the measured system, even though this inference has some uncertainty connected to it. 

From a strict philosophical view, we never observe anything but 
our sensations or perceptions, i.e. the consciousness of sensations,   
that are given to us in an immediate way, according to Kant \cite{Kant}, who stated: 
``Things in space and time are given only in so far as they are perceptions (that is, representations accompanied by sensation)--therefore only through empirical representation''. Thus, in principle, we should follow the observed system, and then the measuring apparatus that interacts with it, hence 
the visible electromagnetic fields that interact with the former, and then 
our retinas that interact with the e.m. field, etc., in a chain with a blurred end due to ill-defined terms 
such as ``consciousness'' and ``sensation''. As von Neumann noted \cite{vonNeumann1932},   
in quantum mechanics there is no need to go up this chain. 
Sometimes, it is sufficient to apply the rules of quantum mechanics directly to the system. This turns out to 
be the case of a projective measurement. Most of the other times, it is necessary to go up just one step, 
by describing both the system and the detecting apparatus. 
Consistency requires that the former case is a limiting case of the latter. It is important to remark that in this second case 
we can not make definite inferences about the system, but we have to leave some 
room for uncertainties. These uncertainties may be classical, in the sense that we do not know the exact state of the 
detector, that the readout has always some noise, etc. But they can also be quantum, if the detector is prepared in a 
superposition. 

Observing the consequences of the quantum nature of the detector, however, is not immediate. 
It can be achieved by making a second observation on the measured system and by making the interaction strength 
weak compared to the coherence scale of the detector in the readout basis. 
When this regime is achieved, weak measurements were demonstrated to allow sequential or 
joint non-projective measurements \cite{Arthurs1965,DiLorenzo2004,DiLorenzo2006,Wei2008,DiLorenzo2011a}, 
to provide new techniques for performing quantum state tomography \cite{Lundeen2011,Lundeen2012,DiLorenzo2013a,Fischbach2012,Wu2013,DiLorenzo2013f,Salvail2013}, 
to provide amplification overcoming technical noise \cite{Hosten2008,Dixon2009,Feizpour2011}, 
and to possibly lead to the disembodiment of dynamical properties from their physical vehicles \cite{Aharonov2012,DiLorenzo2012f}.

\emph{Background.\textemdash}
The weak measurement formalism \cite{Aharonov1988,Duck1989} 
is but perturbation theory applied to the composite system formed by the measured 
system and by a second quantum system, the detector. 
The perturbation is done in the interaction between the two subsystems, while the evolution due to the free Hamiltonians is  considered to be easily computable.  
Furthermore, contrary to textbook perturbation theory, the formalism is applied to conditional probabilities, after a postselection 
on the system is done. As observable probabilities are nonnegative-definite, particular care should be taken while making the 
perturbative expansion, in order to preserve this essential property. In particular, as we show below, the common knowledge 
that all terms of a given order are to be grouped together does not apply. Instead, we shall make an expansion 
in terms of rational functions of the perturbation parameter. 

Usually, in the theory of weak measurement, it is assumed that 
(1) an instantaneous interaction occurs between 
the system and the probe, $H_\mathrm{int}= -\lambda\delta(t)  \Hat{A}\Hat{X}$, where $\Hat{A}$ 
is the variable pertaining to the system 
that is being measured, while $\Hat{X}$ is an operator on the probe; 
(2) the probe is observed through a projective measurement, the projectors $|R\rangle\langle R|$ being the eigenstates 
of an operator $\Hat{R}$, the readout variable; 
(3) the readout variable of the probe, $\Hat{R}$, is canonically conjugated to $\Hat{X}$; 
(4) the initial state of the probe is a pure Gaussian state with a large spread in $R$, $\Delta R\gg \lambda$. 

Here, we shall consider a general interaction of the form 
\begin{equation}
H_\mathrm{int}= -\hbar \lambda g(t)  \Hat{A}\Hat{X},
\end{equation}
where $\int\!dt\, g(t) =1$ and the function $g$ vanishes outside a finite time window $[0,\tau]$. 
In previous works \cite{DiLorenzo2008,DiLorenzo2012e}, we assumed that both $\Hat{A}$ and $\Hat{R}$ 
were conserved. Here, we shall drop this hypothesis as well. Furthermore, we assume no relation between $\Hat{R}$ and 
$\Hat{X}$, allow for an arbitrary initial state of the probe, and consider a nonprojective measurement on the 
probe, characterized by a family of positive operators $\Hat{F}_R$.   
Thus, none among hypotheses (1)-(4) is assumed. While the previous literature has sometimes dropped one or more among 
hypotheses (2)-(4), hypothesis (1) has been always maintained with the exception of Refs.~\cite{DiLorenzo2008,DiLorenzo2012e}. 
The present generalization is particularly interesting in prospective solid state realizations of weak measurement
\cite{Williams2008,Romito2008,Shpitalnik2008,Ashhab2009a,Ashhab2009b,Bednorz2010,Zilberberg2011,Thomas2012}. 

Finally, it is essential, in order to reveal the quantum nature of the probe, that a postselection on the system is made. 
Thus, after the preparation and the interaction, two measurements are made: one on the detector, giving an 
output $R$, and one on the system, giving an output $f$. The quantity of interest is usually (but not always \cite{DiLorenzo2012a}) the average value $\langle R\rangle_f$ of $R$ for a fixed value $f$ of the postselecting measurement. 
To each value $f$ (resp., $R$) is associated a state $\Hat{E}_f$ ($\Hat{F}_R$) of the system (detector). 
The states $\Hat{E}_f, \Hat{F}_R$ are projectors when the measurement is 
a projective one. In general, the measurement is a positive operator valued measure (POVM) \cite{Wiseman2002},  
and the $\Hat{E}_f, \Hat{F}_R$ are not projectors; furthermore, while they are positive definite, they differ from a density operator in that their trace is not necessarily unity. 
Instead, the normalization of the probability 
requires that $\int d\mu_S(f)\Hat{E}_f=\id$ and $\int d\mu_D(R)\Hat{F}_R=\id$, 
 with $\mu_S(f)$ and $\mu_D(R)$ Lebesgues-Stieltjes measures.
Finally, we note that the outputs $R$ are not necessarily real numbers, they can be a set of abstract descriptions 
about the detector, or of our perception thereof. E.g. ``the pointer bounced up and down'', etc. 
However, in most cases, a real number $R$ can be 
associated to each description. Then, it is possible to define the operator 
\begin{equation}\Hat{R}=\int d\mu(R) R \Hat{F}_R.
\end{equation}

Alternatively, one may consider a projective postselection with a probabilistic filtering 
of the data \cite{DiLorenzo2012a}.
We note a difference between the two procedures: 
in the probabilistic based postselection, the state of the system before and after the postselection is 
$\rho_f =\Hat{E}_f/\Tr(\Hat{E}_f)$, in the sense that it optimizes predictions and retrodictions over future and past measurements of the system, respectively;
in the POVM based postselection, instead, the state $\Hat{E}_f$ describes the system in the past, i.e., it is the 
optimal state for retrodiction \cite{Barnett2001}, while the state of the system immediately after the postselection is not 
$\Hat{E}_f/\Tr(\Hat{E}_f)$, but it is $\rho_f = \mathcal{I}_f(\rho'_i)/\Tr[\mathcal{I}(\rho'_i)]$, 
where $\rho'_i$ is the reduced density 
matrix of the system, which differs from the initial $\rho_i$ because of the interaction with the detector, immediately before the postselection, and $\mathcal{I}_f$ is a linear operation on it \cite{Davies1970,Davies1976}, which depends on the details of the POVM; the operator $\Hat{E}_f$ is defined uniquely by $\Tr[\mathcal{I}_f(\rho)]=\Tr[\Hat{E}_f\rho], \,\forall \rho$. 

\emph{Results.\textemdash}
We can now state our main results: 
The conditional average output is 
\begin{widetext}
\begin{align}
&\langle R\rangle_{|f} \simeq
\frac{
%{1}{D}
%\biggl\{
\langle\Hat{R}(\tau)\rangle_0
\!+\int\!\!dt\left( iA'_w(t)\langle[\Hat{R}(\tau),\Hat{X}(t)]\rangle_0\!-\! 
A''_w(t)\langle\{\Hat{R}(\tau),\Hat{X}(t)\}\rangle_0\right)
%\nonumber
%\\
%&\qquad\qquad\qquad\qquad
\!+\!\int \!dt dt'  B_w(t,t') 
\langle\Hat{X}(t')\Hat{R}(\tau)\Hat{X}(t)\rangle_0 
%\biggr\}
}{1-2\int\!\! dt\,  \langle\Hat{X}(t)\rangle_0 A''_w(t) 
+\!\int\!\! dt dt' \langle\Hat{X}(t')\Hat{X}(t)\rangle_0 B_w(t,t')}
\label{eq:main}
,
%\\
%D=&\left\{1-2\lambda\!\!\!\int\!\! dt\,  \langle\Hat{X}(t)\rangle_0 A''_w(t) 
%+\lambda^2\!\iint\!\! dt dt' \langle\Hat{X}(t')\Hat{X}(t)\rangle_0 B_w(t,t')\right\}^{-1}
\end{align}
\end{widetext}
where the symbol $\langle \dots\rangle_0$ denotes averaging over the initial state of the detector,  
the operators defined in the interaction representation, $\Hat{O}(t)= U_t^{(0)\dagger} \Hat{O}(0) U^{(0)}_t $, 
$U^{(0)}$ being the time-evolution operator generated by the non-interacting Hamiltonian $H_0=H_S+H_D$, and 
we introduced the time-dependent weak values 
\begin{align}
&A_w(t)\equiv A'_w(t)+iA''_w(t)=  \lambda g(t) \frac{\Tr_S[\Hat{E}_f(-\tau)\Hat{A}(t)\rho_i]}{\Tr_S[\Hat{E}_f(-\tau)\rho_i]},
\label{eq:aw}
\\
&B_w(t,t')=  \lambda^2 g(t) g(t')\frac{\Tr_S[\Hat{E}_f(-\tau)\Hat{A}(t)\rho_i\Hat{A}(t')]}{\Tr_S[\Hat{E}_f(-\tau)\rho_i]}.
\label{eq:bw}
\end{align}
We remark that as $\Hat{E}_f$ and $\Hat{F}_R$ represent states, not observables, they evolve with the 
opposite time propagator, hence the $-\tau$ argument. 
Notice that for an instantaneous interaction, $g(t)=\delta(t)$, $A_w(t)\to \lambda \delta(t) A_w$, with $A_w$ the standard weak value. 

In the linear regime, when both the first and second order-term in the first factor 
of equation \eqref{eq:main} are negligible compared to 1, the expression reduces to the modified Kubo formula
\begin{align}
\langle R\rangle_f \simeq& \langle\Hat{R}(\tau)\rangle_0+ i\int\!\!dt\,  
\langle\Hat{W}(t)\Hat{R}(\tau)-\Hat{R}(\tau)\Hat{W}^\dagger(t)\rangle_0,
\end{align}
with the non-Hermitian force term
\begin{equation}
\Hat{W}(t) =  A_w(t) [\Hat{X}(t)-\langle\Hat{X}(t)\rangle_0] .
\end{equation}
Notice that if no postselection is made, then $\Hat{E}_f=\id$, the weak value $A_w(t)=\Tr[\Hat{A}(t)\rho_i]$ 
is an average and therefore real, so that  the ordinary Kubo formula \cite{Kubo1957a} is recovered.

\emph{Proof.\textemdash}
The time-evolution operator is 
\begin{equation}
\U =\U_\tau^{(0)}\, \mathcal{T}\!\left\{\exp{\left[i \lambda \int dt\, g(t) \Hat{A}(t) \Hat{X}(t)\right]}\right\} ,
\end{equation}
with $\mathcal{T}$ time-ordering, $\U^{(0)}_\tau = \U^{(S)}_\tau\otimes \U^{(D)}_\tau$ 
free evolution of system and detector, 
while $\Hat{A}(t)=\U_t^{(S)\dagger} \Hat{A} \U_t^{(S)}$ and $\Hat{X}(t)=\U_t^{(D)\dagger} \Hat{X} \U_t^{(D)}$ are the operators in the interaction representation. 
We make a controlled expansion to ``first order'' of the joint probability of observing an output $f$ when making a 
second measurement on the system, and an output $R$ when observing the detector: 
\begin{widetext}
\begin{align}
\P(R,f)\simeq& \frac{1}{N(\lambda)}\Tr\biggl\{\U^{(0)\dagger}_\tau\left(\Hat{E}_f\otimes \Hat{F}_R\right) \U^{(0)}_\tau 
\left[1+i\lambda \int dt\,g(t)\Hat{A}(t) \Hat{X}(t)\right]
%\nonumber\\
%&\qquad\qquad\qquad\qquad\times
(\rho_i\otimes\rho_0) 
\left[1-i\lambda \int dt\,g(t)\Hat{A}(t) \Hat{X}(t) \right] \biggr\}
\nonumber
\\
\simeq& \frac{\Tr_S[\Hat{E}_f(-\tau)\rho_i]}{N(\lambda)}\biggl\{ 
\Tr_D[\Hat{F}_R(-\tau)\rho_0]
+\left[i\int dt\, A_w(t) \Tr_D[\Hat{F}_R(-\tau)\Hat{X}(t)\rho_0]+c.c \right]
\nonumber
\\
&\quad\qquad\qquad\qquad \qquad\qquad\quad 
+\!\iint dt dt' 
B_w(t,t')\Tr_D[\Hat{F}_R(-\tau)\Hat{X}(t)\rho_0\Hat{X}(t')]
\biggr\},
\end{align}
\end{widetext}
with $N(\lambda)$ a normalization. 
We put $\Hat{F}_R(-\tau) = \U^{(D)\dagger}_\tau \Hat{F}_R\U^{(D)}_\tau $ and 
$\Hat{E}_f(-\tau)=\U^{(S)\dagger}_\tau \Hat{E}_f \U^{(S)}_\tau$ the states $\Hat{F}_R$ and $\Hat{E}_f$, respectively, propagated backwards in time to the beginning of the interaction, and defined the time-dependent weak values 
in eqs. \eqref{eq:aw} and \eqref{eq:bw}. 
Notice that, in the present approximation, $N(\lambda)$ is a second-order polynomial in $\lambda$. Thus, while we are making a 
perturbative expansion, we are not approximating the probability by means of a polynomial, as usual, but 
by means of a rational function $f(\lambda)=Q(\lambda)/N(\lambda)$. This guarantees that the probability is positive definite, 
while a naive Taylor expansion may lead to negative probabilities. 

Finally, the probability of postselection is obtained by integrating over $\mu(R)$, yielding 
\begin{align}
\P(f)\simeq& \frac{\Tr_S[\Hat{E}_f(-\tau)\rho_i]}{N(\lambda)}
\biggl\{ 
1-2\int dt\, \langle\Hat{X}(t)\rangle_0 A''_w(t) 
+\iint dt dt'\langle\Hat{X}(t')\Hat{X}(t)\rangle_0 B_w(t,t')
\biggr\}.
\end{align}
The main result \eqref{eq:main} then follows after applying Bayes' rule \cite{Bayes1763} $\P(R|f)=\P(R,f)/\P(f)$ 
and integrating $R \P(R|f)$ over $d\mu(R)$. 

\begin{acknowledgments}
 I thank J. C. Egues for discussions and hospitality at the Instituto de F\'{\i}sica de S\~{a}o Carlos, Universidade de S\~{a}o Paulo. 
This work was performed as part of the Brazilian Instituto Nacional de Ci\^{e}ncia e
Tecnologia para a Informa\c{c}\~{a}o Qu\^{a}ntica (INCT--IQ) and 
it was supported by Funda\c{c}\~{a}o de Amparo \`{a} Pesquisa do 
Estado de Minas Gerais through Process No. APQ-02804-10 and 
by the Conselho Nacional de Desenvolvimento Cient\'{\i}fico e Tecnol\'{o}gico (CNPq) 
through process no. 245952/2012-8. 
\end{acknowledgments}

\bibliography{../weakmeasbiblio}

%%%
%%
%
\end{document}